\documentclass{PoS}
\usepackage{mathtools}
\DeclareMathAlphabet{\mathcal}{OMS}{cmsy}{m}{n}
\newcommand{\pt}{\ensuremath{p_{\mathrm{T}}}}
\newcommand{\ETm}{\ensuremath{E_{\mathrm{T}}^{\text{miss}}}}
\newcommand{\MET}{\ETm}
\newcommand{\cmsSymbolFace}{\mathrm}
\newcommand{\cPqb}{\ensuremath{\cmsSymbolFace{b}}} 
\newcommand{\cPaqb}{\ensuremath{\overline{\cmsSymbolFace{b}}}} 
\newcommand{\bbbar}{\ensuremath{\cPqb\cPaqb}}

\title{Searches for beyond Standard Model Higgs bosons at CMS}

\ShortTitle{Searches for BSM Higgs bosons at CMS}

\author{\speaker{Aruna Kumar Nayak}\thanks{On behalf of CMS Collaboration}\\
        IRFU/SPP, CEA, Saclay, France\\
        E-mail: \email{Aruna.Nayak@cern.ch}}

\abstract{
A search for neutral Higgs bosons in the minimal supersymmetric extension of the Standard Model (MSSM) decaying to a pair of b quarks or a pair of $\tau$ leptons, using events recorded by the CMS experiment at the LHC in 2011 and 2012 at the centre-of-mass energy of 7 TeV and 8 TeV respectively, is presented. 
The result is also presented for a search for the charged Higgs boson that can be produced in the top quark decay with subsequent decay of $H^{+}$ in $\tau^{+}\nu_{\tau}$.
Results are also reported from a search for non-standard-model Higgs boson decays to pairs of new light bosons, each of which decays into the $\mu^{+}\mu^{-}$ final state.
}

\FullConference{The European Physical Society Conference on High Energy Physics -EPS-HEP2013\\
		18-24 July 2013\\
		Stockholm, Sweden}

\begin{document}

\section{Introduction}
A Higgs boson has been recently discovered with a mass around 125 GeV \cite{atlas:Higgs,cms:Higgs} and properties consistent with those expected from the Standard Model.  
However, the exact properties of the newly observed boson still need further investigation. 
Moreover, the SM Higgs boson suffers from quadratically divergent self-energy corrections at high-energy. 
Several extensions of the SM have been proposed to address these divergences. 
Supersymmetry is a well known extension of the SM which allows the cancellation of this divergence. 

The Higgs sector of the Minimal Supersymmetric Standard Model (MSSM) has two scalar doublets which results in five physical Higgs bosons: a light and heavy CP-even h and H, the CP-odd A and the charged boson $H^{\pm}$.
At lowest order the Higgs sector can be expressed in terms of two parameters which are usually chosen as the mass of CP-odd boson ($M_{A}$) and the the ratio of the vacuum expectation values of the two Higgs doublets ($\tan\beta$).
The dominant production mechanisms of the neutral MSSM Higgs boson are gluon fusion and associated production with b quarks. 
The $h,H/A~\rightarrow~\tau^{+}\tau^{-},\bbbar$ constitute the major decay channels for most values of the tan$\beta$ and $M_{A}$. 
The charged Higgs boson can be produced via the decay of the top quark if it is lighter than the top quark.   
For large values of $\tan\beta$, the charged Higgs boson decays to a $\tau$ and a neutrino with $\sim$100\% branching fraction. 
The CMS experiment at the LHC \cite{JINST:3:S08004} has performed searches for MSSM neutral Higgs bosons in both $\tau^{+}\tau^{-}$ and $\bbbar$ decay channels and light charged Higgs boson in $\tau\nu$ final states. 
The latest results on these searches with proton-proton collisions at centre-of-mass energies of 7 TeV and 8 TeV are presented. 

This article also presents the result of a search for the non-Standard Model decay modes of the Higgs boson ($h$), which includes the production of two new light bosons (a), each of which subsequently decay to boosted pairs of oppositely charged muons isolated from the rest of the event activity: $h~\rightarrow~2{\rm a}~+~X~\rightarrow~4\mu~+~X$.
Such a signature can occur in models like the next-to-minimal supersymmetric standard model (NMSSM) and supersymmetric models with additional ``hidden" or ``dark" sectors (Dark SUSY).    
However, the search described in this article is designed to be independent of the details of the specific models, and the results can be interpreted in the context of other models predicting the production of the same final states. 

\section{Search for MSSM $\Phi~\rightarrow~{\boldmath \bbbar}$}
For tan$\beta$ larger than unity, the Higgs field couplings to down-type particles are enhanced by a factor of tan$\beta$ leading to the increase in cross section for Higgs boson production in association with b quarks. Moreover, the decay into b quarks has a very high branching fraction ($\sim$90\%), even at large values of the Higgs boson mass $M_{A}$. 
The result is presented for a search for MSSM neutral Higgs bosons produced in association with at least one b quark, and decaying into a pair of b quarks. 
This analysis is performed using 2.7-4.8 fb$^{-1}$ of data at centre-of-mass energy of 7 TeV collected in 2011 by CMS detector at LHC. 

The dominant background is the production of heavy flavour multijet events containing either three b jets, or two b jets plus a third jet originating from either a charm or a light-flavor parton, which is misidentified as a b jet. 
A signal is searched for in final states characterized either purely by jets (``all-hadronic") or with an additional non-isolated muon (``semileptonic").
The common analysis strategy is to search, in events identified as having at least three b jets, for a peak in the invariant mass distribution of the two leading b jets over the large multijet background.

Events are selected by specialized triggers that include online algorithms for the identification of b jets to tackle the large multijet production at the LHC.
In case of all-hadronic signature, the events are accepted if either two or three jets are produced in the pseudorapidity range |$\eta$| < 2.6 and have $\pt$ above certain thresholds, out of which at least two pass the online b-tagging criteria. 
In the semileptonic analysis, the events were selected at the trigger requiring a muon and the presence of one or two central jets, out of which one or two jets are required to be b-tagged. 
The offline event selection requires to have at least three reconstructed jets within central detector region and with varying $\pt$ thresholds.  
All three leading jets are required to pass a b-tagging selection requirement using the combined secondary vertex algorithm described in \cite{HIG_12_033}, consistent with the online b-tagging demand. 
In the semileptonic analysis, events are required to have a muon that must be contained in one of the two leading jets.
The dominant QCD $\bbbar$ background is estimated using various methods discussed in \cite{HIG_12_033}. 
The signal is extracted, in the all-hadronic analysis, by fitting a linear combination of signal and background templates to the observed histogram in the two-dimensional space of dijet mass and an event b-tag variable constructed from the invariant mass of the secondary vertex in the jets in each event as discussed in \cite{HIG_12_033}, while in case of semileptonic analysis a binned likelihood fit to the invariant mass distribution of the two leading jets is performed.    

\begin{figure}[htp] 
\centering 
\begin{tabular}{c} 
\includegraphics[width=0.4\textwidth]{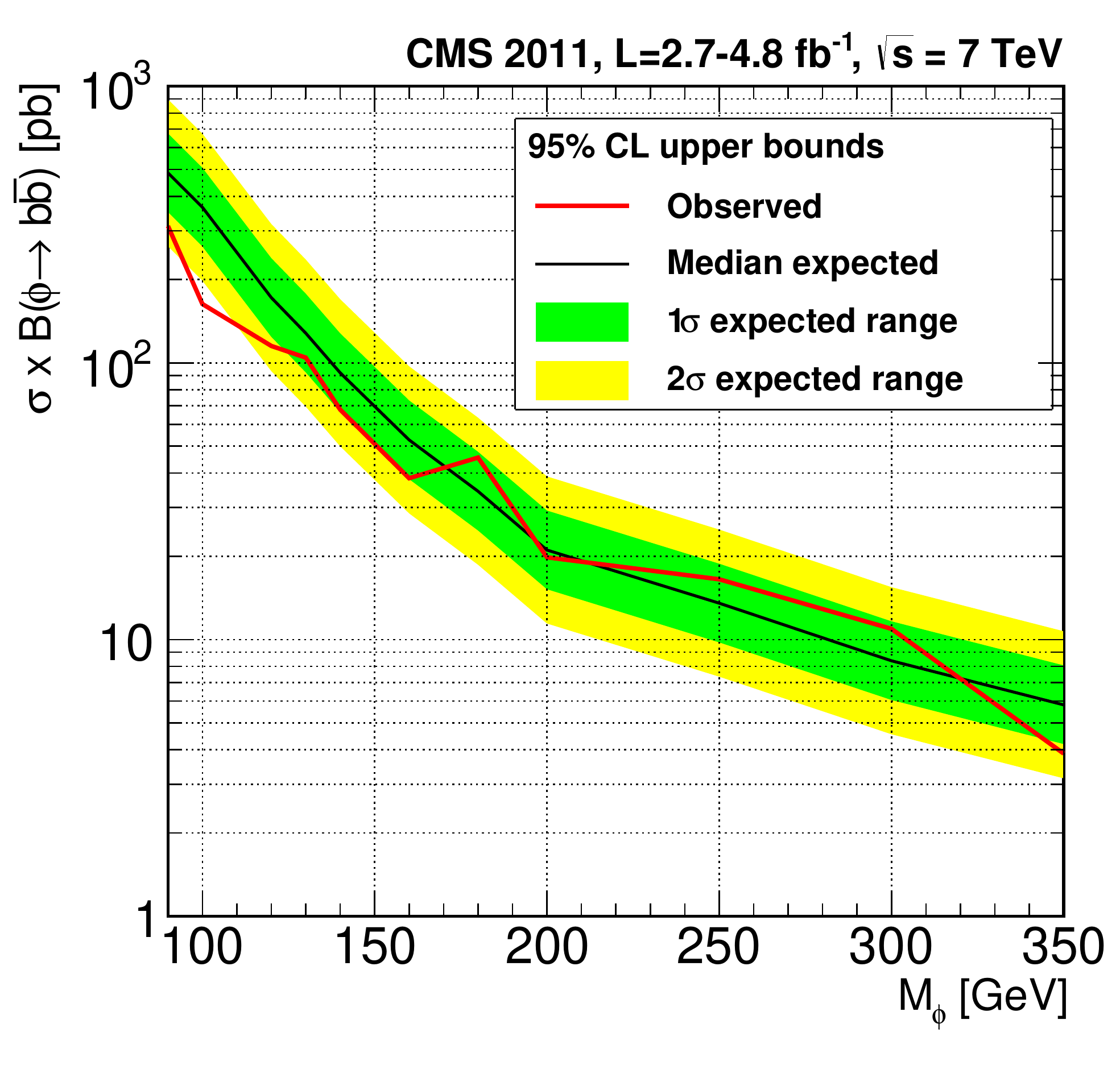} \hfill 
\includegraphics[width=0.4\textwidth]{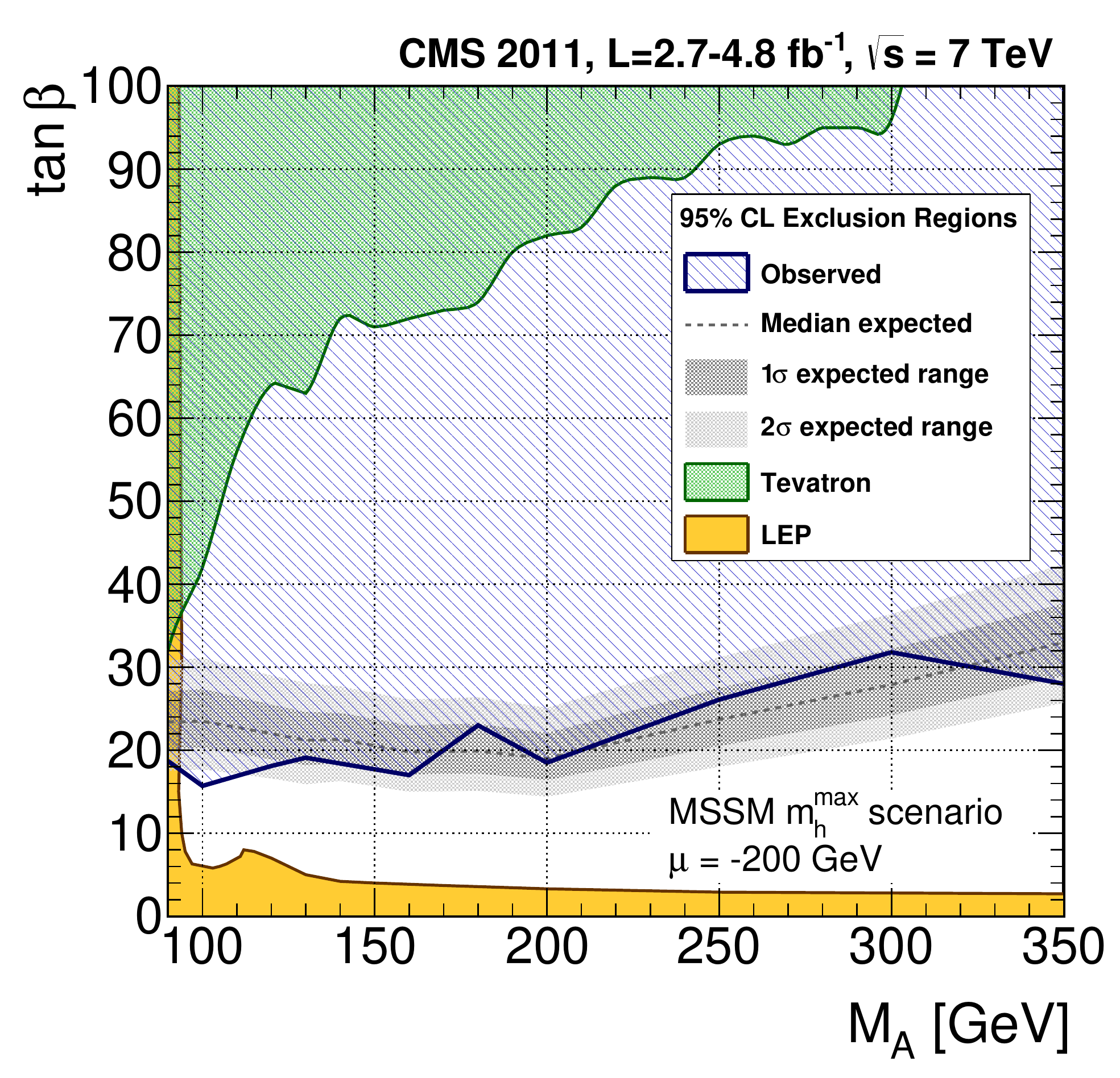}  
\end{tabular} 
\caption{Left: Observed and expected upper limits for the cross section times branching fraction at 95\% CL including statistical and systematic uncertainties for the combined all-hadronic and semileptonic results. Right: Observed upper limits at 95\% CL on tan$\beta$ as function of $M_{A}$ in the $m_{h}^{max}$ benchmark scenario with $\mu$ = -200 GeV. 
} 
\label{fig:bbLimit} 
\end{figure}

No significant deviation from background is observed in either analysis, and the CLs method is used to combine both the results and determine 95\% CL limit on the signal contribution in the data. 
Figure~\ref{fig:bbLimit} (Left) shows the upper limit on the cross section times branching fraction as function of Higgs boson mass $M_{\phi}$. The observed and the expected limit agree within the errors, and no indication of signal is seen.
Figure~\ref{fig:bbLimit} (Right) presents the results in the MSSM framework as a function of the MSSM parameters tan$\beta$ and $M_{A}$, combining the individual results of the two analyses, including all the statistical and systematic uncertainties as well as correlations. 
The result of this analysis extends the sensitivity for MSSM searches in the $\phi~\rightarrow~\bbbar$ decay mode to much lower values of tan$\beta$. 

\section{Search for MSSM $\Phi~\rightarrow~\tau^{+}\tau^{-}$}
The $\tau^{+}\tau^{-}$ final state is particularly interesting due to the enhancement of branching ratios to $\tau$s at large $\tan\beta$. Again the relatively less background contribution compared to $\bbbar$ final state makes the di-$\tau$ final state as the most sensitive channel for neutral Higgs boson search at LHC. 
At CMS, four different $\tau^{+}\tau^{-}$ final states are studied where one or two taus decay leptonically e$\tau_{h}$, $\mu\tau_{h}$, e$\mu$ and $\mu\mu$, where $\tau_{h}$ denotes a hadronic decay of a $\tau$.  
The result is presented for data collected in 2011 and 2012 and correspond to an integrated luminosity of 17 fb$^{-1}$, with 4.9 fb$^{-1}$ at 7 TeV and 12.1 fb$^{-1}$ at 8 TeV. 

The events are selected at the trigger level by combining electron, muon and tau trigger objects whose thresholds and the identification criteria were progressively tightened as the LHC instantaneous luminosity increased over the data-taking period.
In the e$\tau_{h}$ and $\mu\tau_{h}$ final states, the events were selected with an electron or a muon and an oppositely charged $\tau_{h}$, where $\tau_{h}$ is reconstructed using the hadron plus strips algorithm \cite{PAS_HIG_12_050}, based on the particle flow event reconstruction method \cite{PAS_HIG_12_050}. 
In the $e\mu$ and $\mu\mu$ final states events with two oppositely charged leptons are selected.
The QCD multijet background is suppressed by requiring the leptons to be isolated. 
The events coming from W+jets background are rejected by requiring the transverse mass of the electron or muon and the missing transverse energy, $\MET$, to be less than 40 GeV, where $\MET$ is defined as the magnitude of the vector sum of the transverse momenta of the particle-flow candidates. 
The sensitivity of the search for Higgs boson is enhanced by splitting the selected event sample into two mutually exclusive categories:
\begin{itemize}
\item{ {\bf B-Tag: } Events are required to have at least one b-tagged jet with $\pt~>~$20 GeV and not more than one jet with $\pt~>~$30 GeV.}

\item{ {\bf No B-Tag: } Events are required to have no b-tagged jets with $\pt~>~$20 GeV.}
\end{itemize}

The B-Tag category is intended to exploit the production of Higgs bosons in association with b quarks which is enhanced in the MSSM while the no B-Tag category is mainly sensitive to the gluon-fusion Higgs production mechanism. 

The major contribution to the background events comes from $Z~\rightarrow~\tau^{+}\tau^{-}$, which is estimated using a sample of $Z~\rightarrow~\mu\mu$ events where the reconstructed muons are replaced by the reconstructed particles from simulated tau decays. 
The normalization for this process is determined from the measurement of $Z~\rightarrow~\mu\mu$ yield in data. 
The QCD multijet background is estimated using the number of observed same-charge tau pair events while the W+jets background is estimated from events with high transverse mass. 
The other small backgrounds, such as $t\bar{t}$, di-boson and $Z~\rightarrow~ee/\mu\mu$ events, are estimated using simulation. 

\begin{figure}[htp] 
\centering 
\begin{tabular}{c} 
\includegraphics[width=0.5\textwidth]{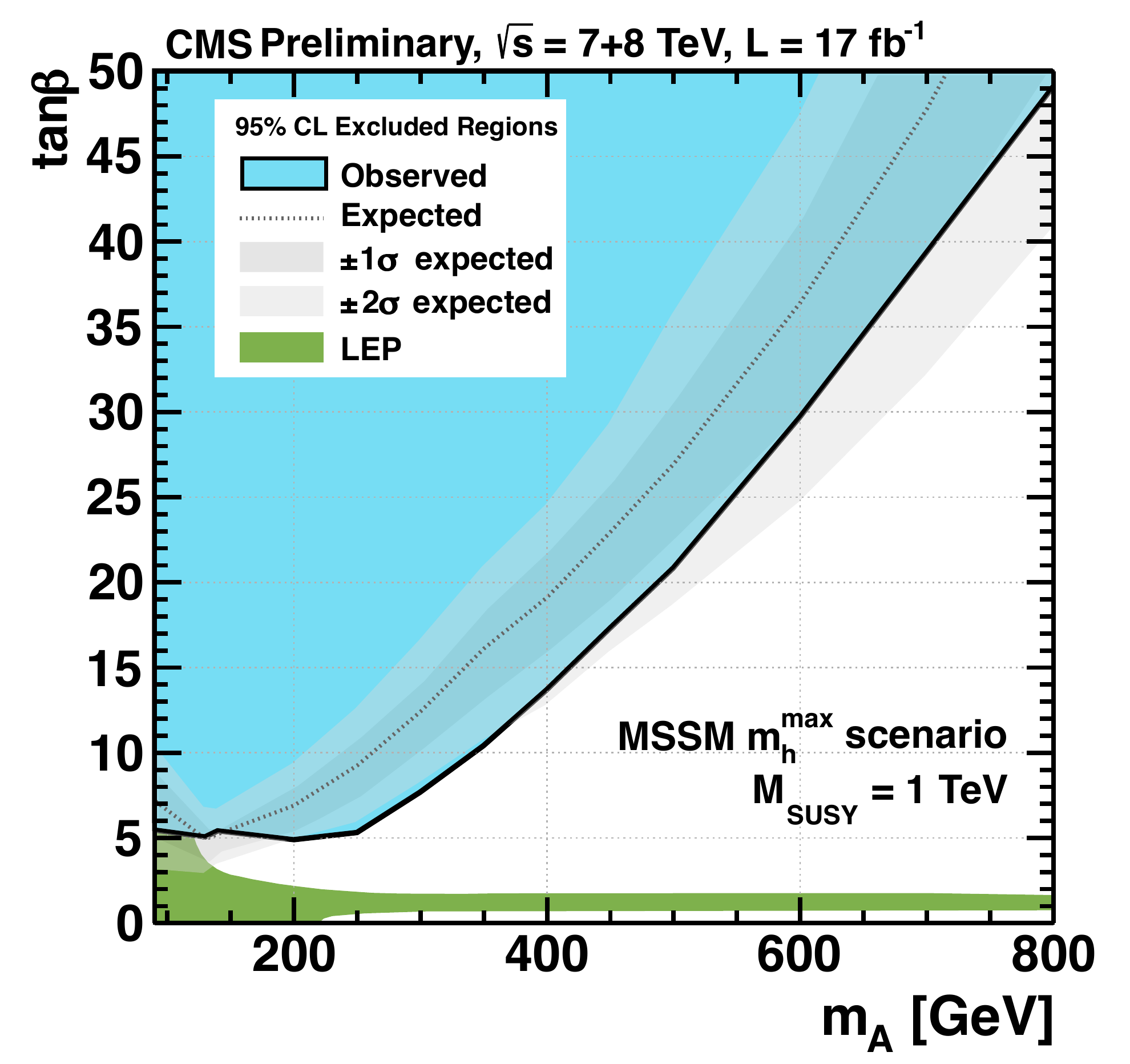}
\end{tabular} 
\caption{Exclusion at 05\% CL in the $\tan\beta$-$M_{A}$ parameter space for MSSM $m_{h}^{max}$ scenario. The exclusion limits from LEP experiments are also shown.} 
\label{fig:tauLimit} 
\end{figure}

To distinguish the signal of Higgs bosons from the background, the tau-pair mass is reconstructed using a maximum likelihood technique \cite{PAS_HIG_12_050}. The algorithm computes the tau-pair mass that is most compatible with the observed momenta of the visible tau decay products and the missing transverse energy reconstructed in the event. 
The algorithm yields a tau-pair mass distribution consistent with the true value and a width of 15-20\%. 
To search for the presence of a Higgs boson signal in the selected events, a binned maximum likelihood fit to the tau-pair invariant mass spectrum is performed. 
Systematic uncertainties are represented by nuisance parameters in the fitting process. 
No evidence for the presence of the Higgs boson signal is found in the invariant mass spectra, therefore 95\% CL upper limit on $\tan\beta$ as function of the pseudoscalar Higgs boson mass $M_{A}$ are set. 
Figure~\ref{fig:tauLimit} shows the 95\% CL exclusion in the $\tan\beta$-$M_{A}$ parameter space for the MSSM $m_{h}^{max}$ scenario. 
The $m_{h}^{max}$ scenario is used since it yields conservative expected limits in the $\tan\beta$-$M_{A}$ plane. 


\section{Search for the charged Higgs boson in top quark decay}
The CMS experiment has performed a search for a light charged Higgs boson produced in the top quark decay \cite{PAS_HIG_12_052, HIG_11_019}. 
The presence of the $t~\rightarrow~H^{+}b,~H^{+}~\rightarrow~\tau^{+}\nu_{\tau}$ decay modes can alter the $\tau$ lepton yield in the decay products of $t\bar{t}$ pairs compared to the Standard Model. 
The search is sensitive to decays of top quark pairs via $t\bar{t}~\rightarrow~H^{\pm}bH^{\mp}\bar{b}$ and $t\bar{t}~\rightarrow~W^{\pm}bH^{\mp}\bar{b}$, where each charged Higgs boson decays into a $\tau$ lepton and a neutrino. 

\begin{figure}[htp]  
\centering  
\begin{tabular}{c}  
\includegraphics[width=0.42\textwidth]{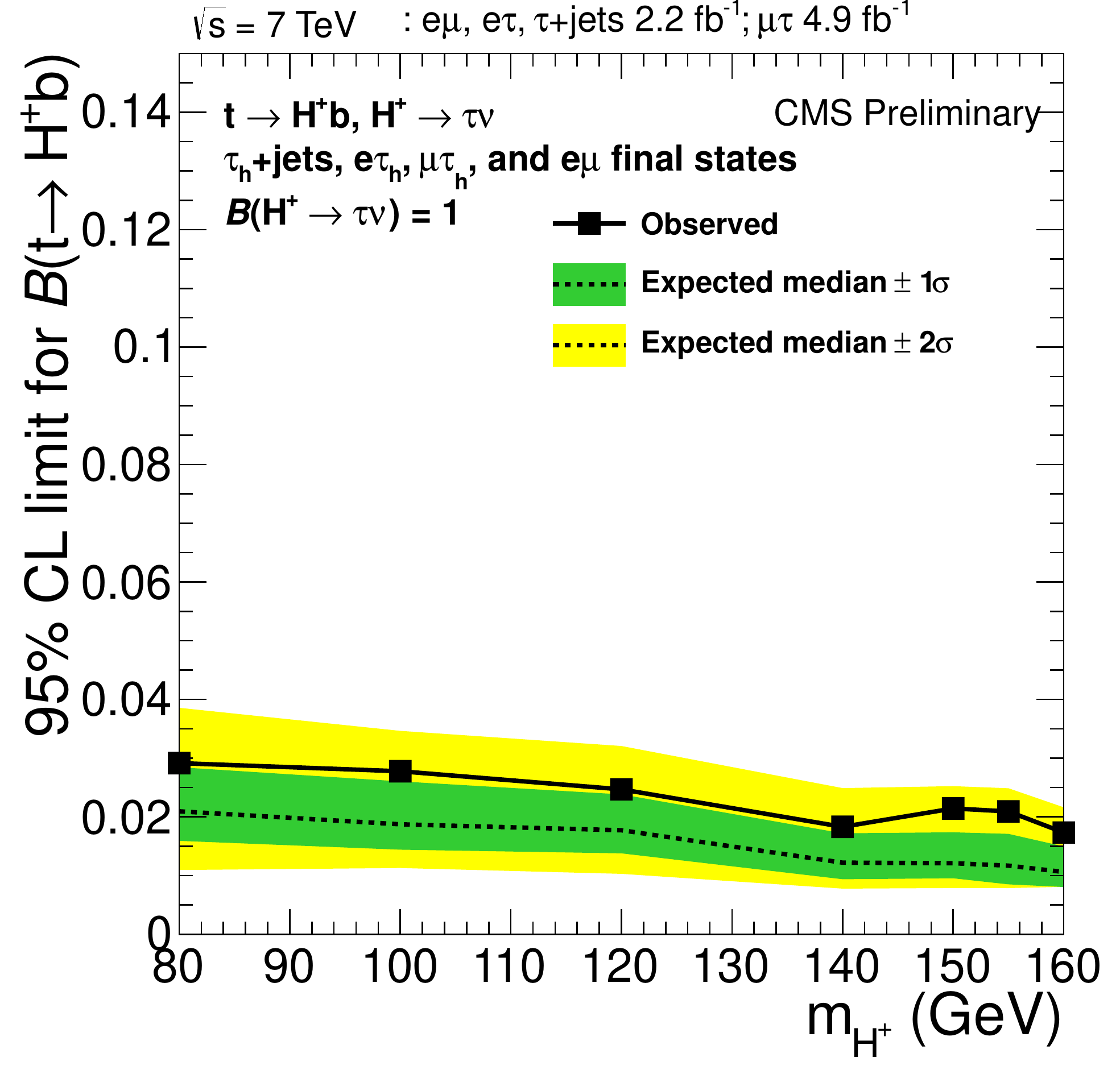} 
\end{tabular}  
\caption{the upper limit on $\mathcal{B}(t~\rightarrow~H^{+}b)$ as a function of $m_{H^{+}}$ obtained from the combination of all final states.}  
\label{fig:HpLimit}  
\end{figure}

Three different final states are studied, all requiring missing transverse energy and multiple jets. 
The first final state involves the production of $\tau_{h}$ and jets ($\tau_{h}$+jets), the second one is where $\tau_{h}$ is produced in association with an electron or a muon ( e$\tau_{h}$ or $\mu\tau_{h}$), and the third one is where an electron and a muon are produced (e$\mu$).  
The analysis uses a data sample recorded in 2011, corresponding to an integrated luminosity of 2.2 to 4.9 fb$^{-1}$ depending on the final state. 
A 95\% CL upper limit is obtained on $\mathcal{B}(t~\rightarrow~H^{+}b)$ assuming $\mathcal{B}(H^{+}~\rightarrow~\tau^{+}\nu_{\tau})$ = 1. 
Figure~\ref{fig:HpLimit} shows the upper limit on $\mathcal{B}(t~\rightarrow~H^{+}b)$ as a function of $m_{H^{+}}$ obtained from the combination of all final states. 

\section{Search for a non-standard-model Higgs boson decaying to a pair of light bosons}
The Higgs sector of the NMSSM consists of 3 CP-even Higgs bosons h$_{1,2,3}$ and 2 CP-odd Higgs bosons a$_{1,2}$. 
The CP-even Higgs bosons h$_{1,2}$ can decay via h$_{1,2}~\rightarrow~$2a$_{1}$, where one of the h$_{1}$ or h$_{2}$ is a SM-like Higgs boson with a mass near 125 GeV/$c^2$ and a$_{1}$ is a new CP-odd light Higgs boson~\cite{PAS_HIG_13_010}. 
The coupling of a$_{1}$ to fermions is proportional to the fermion mass, and can have a substantial branching fraction $\mathcal{B}$(a$_{1}~\rightarrow~\mu^{+}\mu^{-}$) if its mass is within the range $2m_{\mu}~<~m_{{\rm a}_{1}}~<~2m_{\tau}$. 
The generic search presented here looks for two pairs of oppositely charged muons originating from the decay of the Higgs boson~\cite{PAS_HIG_13_010}. 
The analysis is based on a data sample corresponding to an integrated luminosity of 20.65 fb$^{-1}$ of proton-proton collisions at centre-of-mass energy of 8 TeV, obtained in 2012. 

\begin{figure}[htp] 
\centering 
\begin{tabular}{c} 
\includegraphics[width=0.4\textwidth]{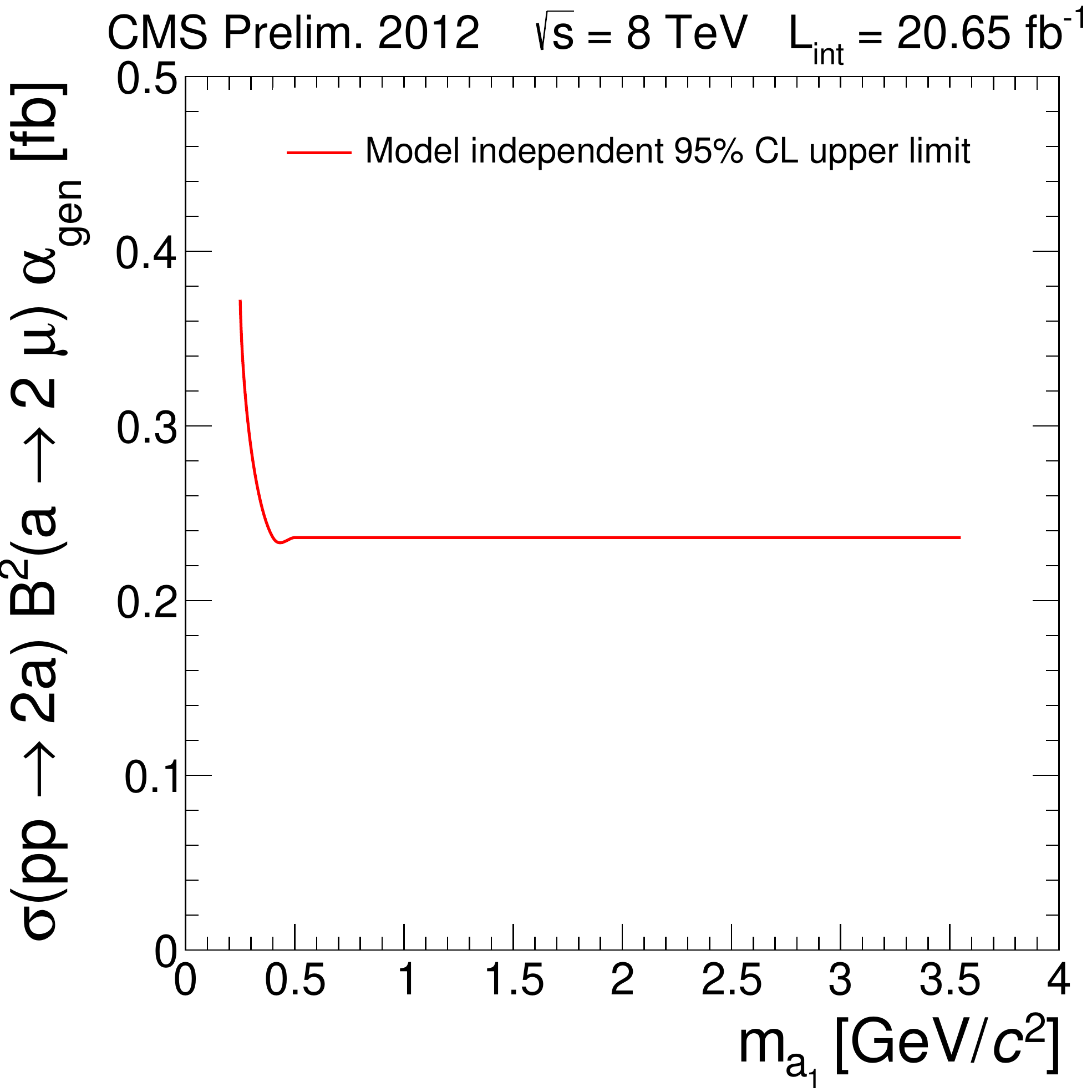} \hfill 
\includegraphics[width=0.4\textwidth]{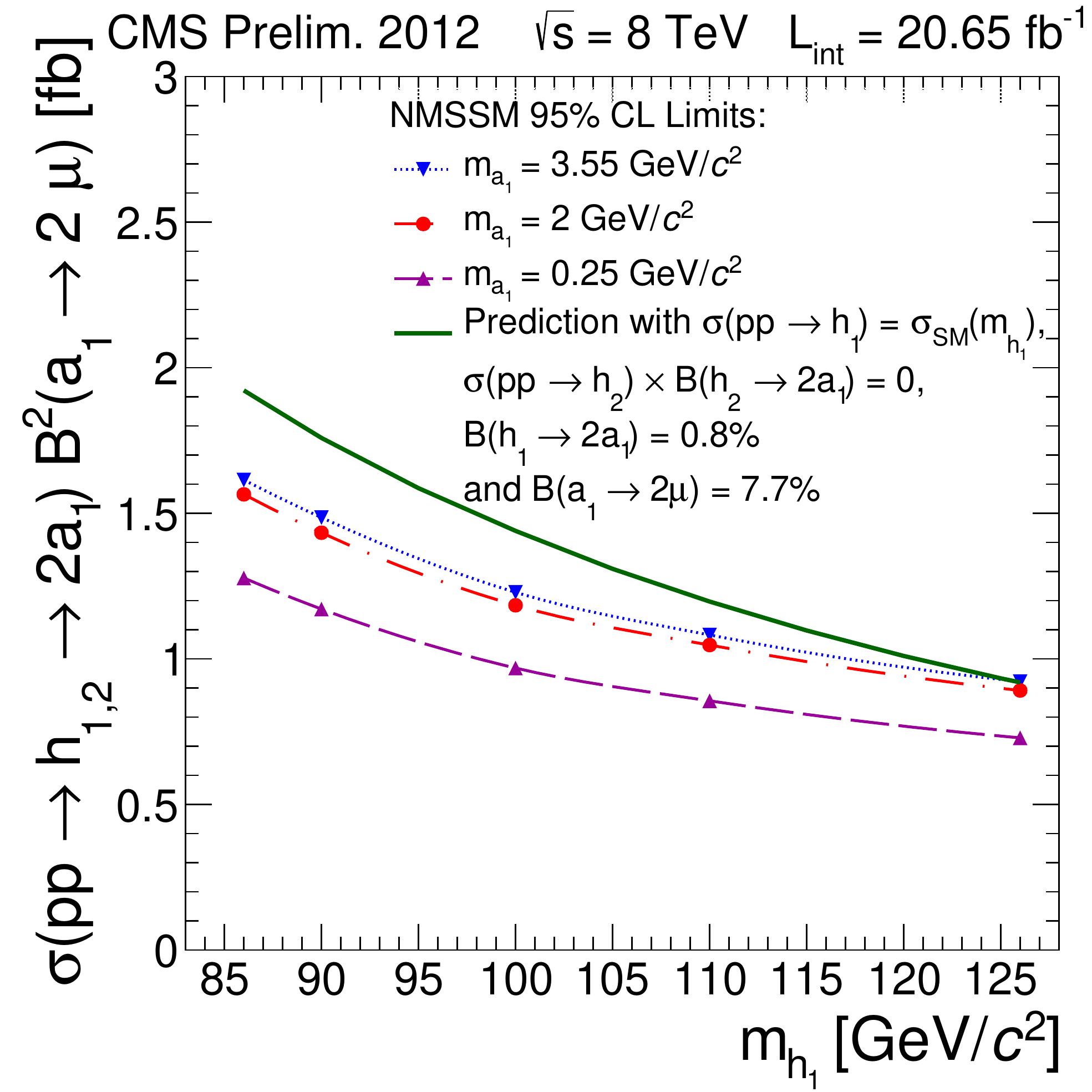}  
\end{tabular} 
\caption{Left: Model independent 95\% CL upper limit on the product of the cross section times branching fraction times acceptance: $\sigma(pp~\rightarrow~2{\rm a}~+X)~\times~\mathcal{B}^{2}({\rm a}~\rightarrow~2\mu)~\times~\alpha_{gen}$.
Right: 95\% CL upper limit as functions of $m_{h_{1}}$, for the NMSSM case.  
} 
\label{fig:nmssmLimit} 
\end{figure}

The events were collected at the trigger level using a dimuon trigger.
The analysis strategy is to select events with exactly two distinct, opposite charged, isolated, muon pairs with dimuon mass less than 5 GeV.
The invariant masses of the two reconstructed dimuon pairs are required to be compatible with each other.
The background contributions are dominated by $\bbbar$ and direct $J/\phi$ pair production events. 
The $\bbbar$ background is estimated from data while the other backgrounds are estimated from simulation.
A 95\% CL upper limit is set on $\sigma(pp~\rightarrow~2{\rm a}~+X)~\times~\mathcal{B}^{2}({\rm a}~\rightarrow~2\mu)~\times~\alpha_{gen}$, where $\alpha_{gen}$ is the generator level kinematic and geometric acceptance of this analysis calculated using generator level information only. 
The efficiency of detector and analysis selection requirements have very weak dependence on the model, thus allowing to set a limit on any arbitrary new physics model predicting similar signature.
The model-independent limit is shown in Fig.~\ref{fig:nmssmLimit} (Left).
The results are interpreted in the context of the NMSSM and the dark-SUSY benchmark models, taking into account the dependence of the signal selection efficiencies on $m_{h}$ and $m_{\rm a}$.
Figure~\ref{fig:nmssmLimit} (Right) shows the exclusion limit in NMSSM model as function of $m_{h_{1}}$ for different values of $m_{{\rm a}_{1}}$. 

\section{Summary}
The CMS experiment has explored the search for the Higgs boson in many promising models beyond SM. 
Searches for the neutral Higgs boson in MSSM decaying to $\bbbar$ and $\tau^{+}\tau^{-}$ are presented. 
No excess of events is observed above the SM background using the data analysed so far, thus setting stringent limit on the $\tan\beta$-$M_{A}$ parameter space. 
The result is presented for the search of a charged Higgs boson in the top quark decay, where the charged Higgs boson decays to a $\tau$ and a neutrino. 
Assuming $\mathcal{B}(H^{+}~\rightarrow~\tau\nu)~=1$, 95\% CL upper limits are set on $\mathcal{B}(t~\rightarrow~H^{+}b)$. 
The result is also presented for the search for a  non-standard-model Higgs boson decays to pair of new light boson, which subsequently decay to pairs of oppositely charged muons.
A model independent upper limit at 95\% CL on the product of the cross section times branching fraction times acceptance is obtained.    
The results have been interpreted in the context of the NMSSM and dark-SUSY models.

\end{document}